\documentclass[letterpaper,superscriptaddress,english,aps,prl,twocolumn, showpacs, amsfonts, amssymb]{revtex4}
\usepackage{epsfig, graphicx,graphics,amsmath,amssymb,float}
\usepackage[T1]{fontenc}
\usepackage[latin9]{inputenc}
\usepackage{amsmath}
\usepackage{amssymb}

\usepackage{amscd}
\usepackage{bm}
\usepackage{psfrag}

\usepackage{babel}

\usepackage{bbm}

\usepackage{color}
\usepackage[bookmarks=true,colorlinks,linkcolor=blue,urlcolor=blue,citecolor=blue]{hyperref}

\newcommand{\be}{\begin{equation}}
\newcommand{\ee}{\end{equation}}
\newcommand{\bea}{\begin{eqnarray}}
\newcommand{\eea}{\end{eqnarray}}

\newcommand{\mc}{\mathcal}

\begin{document}

\title{Chiral spin-orbital liquids with   nodal lines}
\author{W. M. H. Natori}
\affiliation{Instituto de F\'{i}sica de S\~ao Carlos, Universidade de S\~ao Paulo, C.P. 369, S\~ao Carlos, SP,  13560-970, Brazil}
\author{E. C. Andrade}
\affiliation{Instituto de F\'{i}sica de S\~ao Carlos, Universidade de S\~ao Paulo, C.P. 369, S\~ao Carlos, SP,  13560-970, Brazil}
\affiliation{Instituto de F\'{i}sica Te\'{o}rica, Universidade Estadual Paulista,
Rua Dr. Bento Teobaldo Ferraz, 271 - Bloco II, 01140-070, S\~ao Paulo, SP, Brazil}
\author{E. Miranda}
\affiliation{Instituto de F\'{i}sica Gleb Wataghin, Unicamp, Rua S\'{e}rgio Buarque de Holanda, 777, CEP 13083-859 Campinas, SP, Brazil}
\author{R. G. Pereira}
\affiliation{Instituto de F\'{i}sica de S\~ao Carlos, Universidade de S\~ao Paulo, C.P. 369, S\~ao Carlos, SP,  13560-970, Brazil}
\affiliation{
International Institute of Physics, Universidade Federal do Rio Grande do Norte, 59078-970 Natal-RN,  Brazil, and \\
Departamento de F\'isica Te\'orica e Experimental, Universidade Federal do Rio Grande do Norte, 59072-970 Natal-RN, Brazil}


\begin{abstract}
Strongly correlated materials with strong spin-orbit coupling hold  promise for  realizing topological phases with  fractionalized excitations. Here we propose a chiral spin-orbital liquid as a stable phase of a realistic model for heavy-element   double  perovskites. This spin liquid state has  Majorana fermion excitations with a gapless spectrum characterized by  nodal lines along the edges of the Brillouin zone. We show that the nodal lines are topological defects of a non-Abelian Berry connection  and that the system exhibits  dispersing surface states. We discuss some experimental signatures of this state and compare them with properties of the spin liquid candidate   Ba$_{2}$YMoO$_{6}$.
\end{abstract}
\maketitle

Quantum spin liquids (QSLs) are Mott insulators in which   quantum fluctuations prevent 
long-range magnetic order down to zero temperature \cite{Balents2010}.
They have received both experimental and theoretical attention due to predictions 
of unusual phenomena such as spin-gapped phases with  topological order   or  gapless phases without  spontaneous
   breaking of continuous symmetries \cite{Wen2004}. In recent years the evidence for  QSLs in nature has started to look more auspicious thanks to the discovery of new compounds that realize the Heisenberg model on   frustrated   lattices \cite{Han2012}. 
 While  frustration is a desirable ingredient, seminal work by Kitaev \cite{Kitaev2006} has demonstrated that bond-dependent   exchange interactions may provide another route towards   QSL ground states. 
The key idea is that a  spin-$1/2$ model on the (bipartite) honeycomb lattice
with  judiciously chosen anisotropic interactions can be rewritten  in terms of free Majorana fermions hopping in the background of a static Z$_2$ gauge field.  The result is a QSL with exotic fractional excitations. The same idea has been applied to construct other exactly solvable models, including cases of higher spins  \cite{Yao2009,Mandal2009,Wu2009,Hermanns2004,OBrien2016}.

From a broader perspective, Kitaev's model is an instance of a \emph{quantum compass model} \cite{Dorier2005,Lacroix2011,Nussinov2015}. Although   Kitaev-type exactly solvable  models are  artificial,   the kind of anisotropic interactions   they presuppose arises naturally in Mott insulators with orbital degeneracy and strong spin-orbit coupling \cite{Jackeli2009,Chaloupka2010}. There is recent evidence that bond-dependent interactions are  dominant  in  Na$_2$IrO$_3$  \cite{Chun2015}. While   this compound is in a zigzag-ordered phase at low temperatures, the prospect of finding QSLs in compass models suggests inspecting other families of heavy-element transition metal oxides \cite{Pesin2010,WanPRB2011,Witczak2014}. 

All the conditions leading to   quantum compass models
can be found in Mott-insulating rock-salt-ordered double perovskites \cite{Chen2010}. Given the  chemical formula A$_2$BB$^\prime$O$_6$, particularly interesting  properties are found in  compounds where the B$^\prime$ magnetic ions have a 4d$^1$ or 5d$^1$ configuration. These ions are arranged  in a face-centered-cubic (fcc) lattice, which, unlike the honeycomb lattice, \emph{is} geometrically frustrated. The  magnetic properties    within this family are diverse  \cite{Wiebe2002,Stitzer2002,Yamamura2006,Erickson2007}, but the material that stands out is Ba$_2$YMoO$_6$ \cite{Vries2010,Aharen2010,Carlo2011,Vries2013}. Despite a Curie-Weiss temperature   $\Theta_{\text{cw}}\approx -160$ K \cite{Vries2010}, several experiments point to the absence of long-range order 
down to $T\sim2$ K. Moreover, there is no sign of structural transitions, implying that    the lattice remains     cubic    at low temperatures. Thus,   the degeneracy of the  $t_{2g}$ orbitals is only partially lifted by the   spin-orbit coupling, leading to a low-lying $j=3/2$ quadruplet \cite{Vries2010}. The effective  model      contains   bond-dependent interactions between nearest-neighbor $j=3/2$ spins and is closely related to $\Gamma$-matrix generalizations of  Kitaev's model \cite{Yao2009,Wu2009}. Remarkably, the analysis in \cite{Chen2010} revealed  that, when antiferromagnetic exchange is  dominant,     ordered phases become unstable against quantum fluctuations, making this an interesting   point to look for QSLs.

In this Letter we investigate   a QSL in a realistic model for double perovskites with strong spin-orbit coupling. Using a representation of $j=3/2$ operators in terms of six Majorana fermions, we start by showing  that a hidden SU(2) symmetry of the Hamiltonain becomes an explicit SO(3) symmetry for three of these fermions, whereas the other three    exhibit a compass-model-type Z$_3$ symmetry. As the model is not exactly solvable, we proceed with a mean-field approach and propose a  spin liquid ansatz   that preserves the SO(3) and Z$_3$ symmetries.  
The ansatz  breaks  inversion and time reversal symmetry, thus  describing  a \emph{chiral} spin liquid.  Most interestingly, we find that  the excitation spectrum is gapless along  \emph{nodal lines}   which are  vortices of a Berry connection in momentum space. This feature makes this chiral spin liquid a strongly correlated analogue  of   line-node semimetals and superconductors discussed in the context of topological phases of weakly interacting electrons \cite{Burkov2011,Chiu2014,Yang2014,Bian2015} and photonic crystals \cite{Lu2013}. Going beyond the mean-field level, we use variational Monte Carlo (VMC) techniques \cite{Gros1989,Ceperley1977,Edegger2005} to show that our spin liquid state yields a remarkably low energy and should be regarded as a competitive candidate for the ground state of the spin-orbital model. Finally, we argue that  the vanishing density of states at low energies predicted by our theory can account for some unusual properties observed in Ba$_2$YMoO$_6$. 

The  spin-orbital model for cubic double perovskites with d$^1$ electronic configuration is given by \cite{Chen2010}
\begin{equation}
 H=J\underset{\alpha,\langle ij\rangle\in\alpha}{\sum}\left(\textbf{S}_{i,\alpha}\cdot\textbf{S}_{j,\alpha}-\frac{1}{4}n_{i,\alpha}n_{j,\alpha}\right)-\lambda\sum_i\textbf{l}_{i}\cdot\textbf{S}_{i}.\label{eq:Horiginal}
\end{equation}
Here $J>0$ is the antiferromagnetic exchange   between nearest-neighbor spins   and $\lambda>0$ is the atomic spin-orbit coupling.  The index $\alpha$   labels both   planes (XY, YZ or
XZ) and   $t_{2g}$ orbitals ($d_{xy}$, $d_{yz}$ or $d_{xz}$) \cite{Tokura2000}.
The operators   $n_{i,\alpha}$ and  $\textbf{S}_{i,\alpha}$ describe   the number  and the spin  of electrons occupying the $\alpha$  orbital  on site $i$, with the constraint $\sum_\alpha n_{i,\alpha}=1$, and $\mathbf l_i$ is the effective $l=1$ orbital angular momentum of the $t_{2g}$ states \cite{Ballhausen1962}. The total spin on site $i$ is 
 $\mathbf S_i=\sum_\alpha \mathbf S_{i,\alpha }$.  
 
In the  regime    $\lambda\gg J$, spin and orbital operators can be projected into the low-energy  subspace of total angular momentum $j=3/2$ \cite{Chen2010}. The projected Hamiltonian $\tilde H=\mc P_{3/2} H\mc P_{3/2}$, where $\mc P_{3/2}$ is the projector, contains multipolar interactions in terms of  $\mathbf J_i=\mathbf l_i+\mathbf S_i$.  Our  first step is to   introduce operators  $\mathbf s$ and $\boldsymbol  \tau$ at  each site   as  \be
\textbf{s}=\frac{1}{2}(-\Gamma^{23},\Gamma^{13},\Gamma^{12}),\quad 
 \boldsymbol{\tau}=\frac{1}{2}(\Gamma^{4},-\Gamma^{45},\Gamma^{5}).\ee
The notation   refers to  five Dirac $\Gamma$ matrices given explicitly by   
\bea
\Gamma^{1}  &=&\sigma^{z}\otimes\sigma^{y},\;\Gamma^{2}=\sigma^{z}\otimes\sigma^{x},\;\Gamma^{3}=\sigma^{y}\otimes \mathbbm1,\nonumber \\
\Gamma^{4}  &=&\sigma^{x}\otimes\mathbbm1,\;\Gamma^{5}=-\Gamma^{1}\Gamma^{2}\Gamma^{3}\Gamma^{4},
\eea
where $\sigma^a$, $a\in \{x,y,z\}$, are  Pauli matrices, 
and  10  matrices $\Gamma^{\mu \nu}\equiv\left[\Gamma^{\mu},\Gamma^{\nu}\right]/(2i)$  \cite{Murakami2004,Yao2009}.  
 The components of $\mathbf s$ and $\boldsymbol \tau$
satisfy the   SU(2) algebra $[s^a,s^b]=i\epsilon^{abc}s^c$, $[\tau^a,\tau^b]=i\epsilon^{abc}\tau^c$, and $[s^a,\tau^b]=0$. The
relation between the basis  of $J^z$  and the basis 
$|s^{z},\tau^{z}\rangle$, with $s^z,\tau^z\in\{+,-\}$, is $
\left|J^{z} =\pm \frac{3}{2}\right\rangle=|\mp,+\rangle,\left|J^{z}=\pm \frac{1}{2}\right\rangle=|\pm ,-\rangle$. 
%Local operators can be written as linear combinations of the 15 matrices $s^a,\tau^b,s^a\tau^b$.  

In the new representation the projected Hamiltonian assumes  a relatively simple form: \be 
\tilde  H=\frac{J}{9}\underset{\langle i,j\rangle\in\alpha}{\sum}\left(\textbf{s}_{i}\cdot \textbf{s}_{j}-\frac{1}{4}\right)(1-2\tau_{i}^{\alpha})(1-2\tau_{j}^{\alpha}),\label{eq:AFM H}
\ee
where  $\tau^{\alpha}$ are    given by $\tau^{xy}=\tau^{z}$, $\tau^{yz(xz)}={\frac{1}{2}(-\tau^{z}\pm\sqrt{3}\tau^{x})}$.  A few comments are in order. First, Eq. (\ref{eq:AFM H}) has the familiar form of a Kugel-Khomskii model \cite{Kugel1982,Feiner1997}.  However, here the Kugel-Khomskii coupling involves   \emph{pseudospins} $\mathbf s$ and \emph{pseudo-orbitals}  $\boldsymbol \tau$ defined   in the $j=3/2$ subspace, where the original spins and orbitals are highly entangled. Second, the Hamiltonian commutes with $\mathbf s_{\text{tot}}=\sum_i \mathbf s_i$. This is a manifestation of the hidden global SU(2) symmetry discussed in \cite{Chen2010}. This continuous  symmetry is unexpected, given that spin-orbit coupling breaks the conservation of   $\mathbf J_{\text{tot}}=\sum_i\mathbf J_i$, but appears in related  models for $t_{2g}$ orbitals \cite{Harris2003} and   at special points of the Kitaev-Heisenberg model \cite{Chaloupka2015}. Finally, the pseudo-orbital coupling has the form of a 120$^\circ$ compass model   \cite{Nussinov2015}. There is a Z$_3$ symmetry generated by $U_{3}=e^{-i\frac{2\pi}{3}\tau_{\text{tot}}^y}$ followed by a C$_3$ rotation of the $\alpha$ planes. 
%This form of the Hamiltonian clarifies that the mechanism   behind the quantum melting of conventional magnetic order in double perovskites \cite{Chen2010} is twofold: ({\it i}) low-energy fluctuations of pseudospins due to a hidden SU(2) symmetry, and ({\it ii}) frustration in the pseudo-orbital sector due to compass-type interactions. 

In analogy with the spin liquid in Kitaev's model \cite{Kitaev2006}, we now introduce a Majorana parton representation for the generators of SU(4) (i.e. the basis of $j=3/2$ operators). We write $\mathbf s$ and $\boldsymbol \tau$ operators as \cite{tsvelik2007quantum,Coleman1994,Wang2009,Biswas2011,Kopietz2013,Trebst2015}
\be
 s^a_{j}  =-\frac{i}{4}\epsilon^{abc}\eta^b_{j}\eta^c_{j},\qquad\tau^a_{j}=-\frac{i}{4}\epsilon^{abc}\theta_{j}^b\theta^c_{j}.
\ee
The components of  $\boldsymbol \eta_j$ and $\boldsymbol \theta_j$   are Majorana fermions that obey   $\{\gamma_j^a,\gamma_l^b\}=2\delta_{jl}\delta^{ab} $, where $\gamma\in \{\eta,\theta\}$. As the signs of the    fermions can be changed    ($\boldsymbol{\eta}\rightarrow-\boldsymbol{\eta}$,
$\boldsymbol{\theta}\rightarrow-\boldsymbol{\theta}$) without affecting  the physical operators,  this representation bears a
Z$_{2}$ redundancy. To eliminate the extra states, one can impose the local constraint \cite{Wang2009}
\be
D_{j}  \equiv i\eta^1_{j}\eta^2_{j}\eta^3_{j}\theta^1_{j}\theta^2_{j}\theta^3_{j},\qquad D_{j}=1\;\forall j.\label{eq:constraint}
\ee
With this constraint we have $s_{j}^{a}\tau_{j}^{b}=-\frac{i}{4}\eta^a_{j}\theta^b_{ j}$
, and Hamiltonian (\ref{eq:AFM H}) becomes  quartic in Majorana fermions:
\bea
\tilde{H} & =&-\frac{NJ}{6}+\frac{J}{36}\sum_{\langle i,j\rangle\in\alpha}\left[\left(\sum_{a<b} \eta^a_{i}\eta^a_{j}\eta^b_{i}\eta^b_{j}\right)\right.\nonumber \\
 && +(\eta^1_{i}\eta^2_{i}\eta^3_{j}+\eta^2_{i}\eta^3_{i}\eta^1_{j}+\eta^3_{i}\eta^1_{i}\eta^2_{j})\bar{\theta}^\alpha_{j}+( i\leftrightarrow j)\nonumber \\
 && \left.+\bar{\theta}^\alpha_{i}\bar{\theta}^\alpha_{j}\boldsymbol{\eta}_{i}\cdot \boldsymbol{\eta}_{j}-\theta^\alpha_{i}\theta^\alpha_{j}\theta^2_{i}\theta^2_{j}\right],\label{quarticH}
\eea
where $N$ is the number of sites and     $\theta^{\alpha}$ and $\bar{\theta}^{\alpha}$ are defined by
$\theta^{xy}=\theta^1$, $\theta^{yz(xz)}=\frac{1}2(-\theta^1\mp \sqrt3\theta^3)$, and $\bar\theta^{xy}=\theta^3$, $\bar\theta^{yz(xz)}=\frac{1}2(-\theta^3\pm \sqrt3 \theta^1)$.

Hamiltonian (\ref{quarticH}) is invariant under global SO(3) rotations  of the $\boldsymbol\eta$ vector. The couplings involving the components of $\boldsymbol \theta$ 
have only a discrete symmetry, namely the octahedral point group  symmetry O$_{\text{h}}$ of the   lattice. The latter contains the Z$_3$ that rotates   $\theta^\alpha$ and $\bar\theta^\alpha$   by 120$^\circ$ in the $(\theta^1,\theta^3)$ plane. In addition, $\tilde H$  is invariant under time reversal   $T=K e^{-i\pi J_{\text{tot}}^y}$, where $K$ denotes complex conjugation. In terms of Majorana fermions,   $T=K\prod_j\theta_j^1\theta_j^3$.

Next, we perform a mean-field decoupling of Hamiltonian (\ref{quarticH}). This is equivalent to neglecting fluctuations of the Z$_2$ gauge field and   yields qualitatively correct results as long as  the system is in a QSL  phase with deconfined Majorana fermions \cite{tsvelik2007quantum}.  Our choice of mean field ansatz is guided by the condition of preserving the SO(3)  and   Z$_3$ symmetries. This   restricts the set of nonzero  parameters $\langle \gamma^a_i\gamma^b_j\rangle$. We obtain \bea
\tilde H_{\text{MF}}&=&-\frac{NJ}{6}+\frac{J}{36}\underset{\langle j,l\rangle\in\alpha}{\sum}\left[i(2u_{jl}+\bar{w}^\alpha_{jl})\boldsymbol{\eta}_{j}\cdot\boldsymbol{\eta}_{l}\right.\nonumber \\
 && +3iu_{jl}\bar{\theta}^\alpha_{j}\bar{\theta}^\alpha_{l}-iw^\alpha_{jl}\theta^2_{j}\theta^2_{l}-iv_{jl}\theta^\alpha_{j}\theta^\alpha_{l}\nonumber \\
 & &\left.+3u_{jl}^{2}+3\bar{w}^\alpha_{jl}u_{jl}-w^\alpha_{jl}v_{jl}\right],\label{eq:Hmf}
\eea 
where $iu_{jl}=\langle\eta^a_{j}\eta^a_{l}\rangle$, $iv_{jl}=\langle\theta^2_{j}\theta^2_{l}\rangle$, 
$iw^\alpha_{jl}=\langle\theta^\alpha_{j}\theta^\alpha_{l}\rangle$, and $i\bar{w}^\alpha_{jl}=\langle\bar{\theta}^\alpha_{j}\bar{\theta}^\alpha_{l}\rangle$ play the role of imaginary hopping amplitudes. 
Note that the symmetry implies decoupling of $\eta^a$ and $\theta^2$   fermions at the  level of bilinear terms; yet,   $\theta^1$ and $\theta^3$ remain coupled.

\begin{figure}[t]
\begin{center}
\includegraphics*[width=.8\columnwidth]{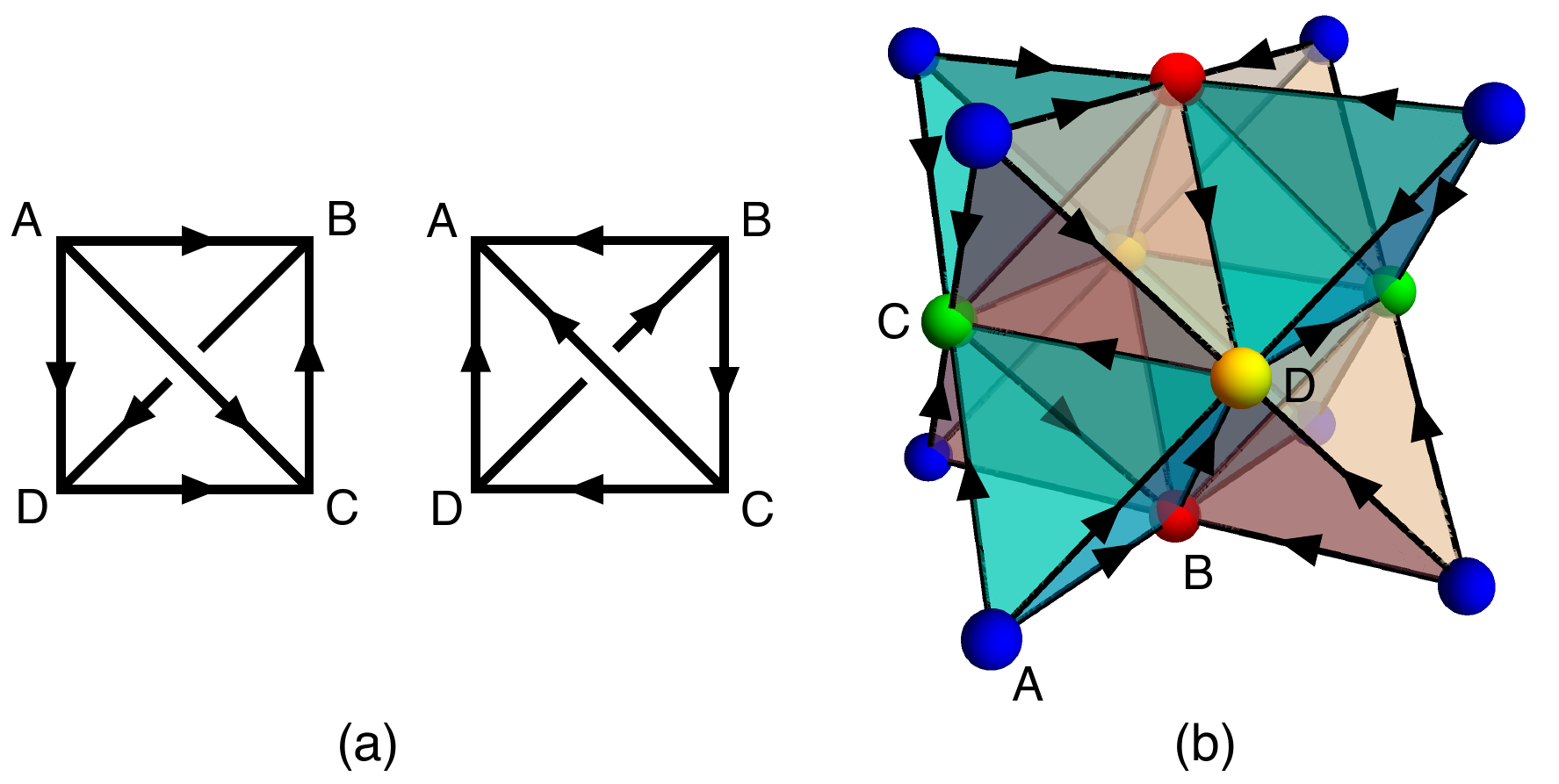}
\end{center}
\caption{\label{fig:Ans=0000E4tze}(color online). (a) 
Two gauge-inequivalent hopping configurations with the same flux of the Z$_2$ gauge field   on all   faces of a tetrahedron.   (b) Four-sublattice ansatz   on the fcc lattice. The sign of the  outward flux  alternates between edge-sharing tetrahedra (represented by different color fillings). 
 }
\end{figure}

Seeking a translationally invariant ansatz, we set the   order parameters  to have uniform magnitude: $u_{ij}=u\sigma^u_{ij}$, $v_{ij}=v\sigma_{ij}^v$, $w^{xy}_{ij\in \text{XY}}=w \sigma^w_{ij} $, $\bar w^{xy}_{ij\in \text{XY}}=\bar w \sigma^{\bar w}_{ij} $,  with $u, v,w,\bar w$ to be determined by self-consistent equations, whereas the $\sigma$'s are chosen to be $\pm1$ on each bond.  Since  e.g. $u_{ij}=-u_{ji}$, the choice of $\sigma^{u}_{ij}$ is equivalent to a choice of bond  orientation and determines the gauge-invariant flux   through elementary plaquettes. Noticing that the fcc lattice can be viewed as a network of edge-sharing tetrahedra, we obtain a symmetric ansatz by requiring that  the Z$_2$ fluxes, e.g. $\chi^{u}_{jkl}\equiv i\sigma^u_{jk}\sigma^u_{kl}\sigma^u_{lj}$, be the same on all faces of a given tetrahedron, with sites $jkl$ on every triangle oriented counterclockwise with respect to an outward normal vector. This leads to the four-sublattice ansatz illustrated in   Fig. \ref{fig:Ans=0000E4tze}.

Let us discuss the symmetry of our   ansatz.  First, we note that the Z$_2$ gauge flux through  triangles is odd under time reversal   and is related to the  spin chirality order parameter \cite{Baskaran1989,Wen1989}.  The state  also breaks inversion  $P$; this can be seen from   Fig. \ref{fig:Ans=0000E4tze}(b), since a mirror-plane reflection exchanges tetrahedra with opposite chiralities. Thus, our ansatz describes a chiral spin liquid with spontaneous breaking of $P$ and $T$.   However,   $PT$ is still a symmetry.  Similarly, a projective symmetry group analysis  \cite{Wen2004} shows that broken rotational  symmetries can be combined with the broken time reversal to restore an O$_{\text h}$ point group symmetry,   ensuring  the orbital degeneracy assumed at the outset  (see Supplemental Material \cite{Suppl}).

Having  fixed the    ansatz, we can calculate the resulting spectrum of  the Majorana fermions. For simplicity, first we focus on the mean-field Hamiltonian for $\gamma\in \{\eta^a,\theta^2\}$, i.e., the flavors which are decoupled in Eq. (\ref{eq:Hmf}). In this case the Hamiltonian can be written in the form \be
 \tilde H_{\text{MF}}=\sum_{\mathbf k\in\frac12\text{BZ}}\gamma^\dagger_{\mathbf k}\mc H (\mathbf k)\gamma^{\phantom\dagger}_{\mathbf k}=|t|\sum_{\mathbf k\in\frac12\text{BZ}}\gamma^\dagger_{\mathbf k}(\mathbf h_{\mathbf k}\cdot \boldsymbol \Sigma)\gamma^{\phantom\dagger}_{\mathbf k},\ee
 where $t=t(u,v,w,\bar w)$ is the corresponding hopping amplitude in Eq. (\ref{eq:Hmf}), $\gamma_{\mathbf k}^\dagger=(\gamma_{\mathbf k \text A}^\dagger,\gamma_{\mathbf k \text B}^\dagger,\gamma_{\mathbf k \text C}^\dagger,\gamma_{\mathbf k \text D}^\dagger)$ is a   spinor with   components labeled by sublattice index, $\mathbf h_{\mathbf k}=4\left(\cos\frac{k_x}{2}\cos\frac{k_y}{2},\cos \frac{k_y}{2}\cos\frac{k_z}{2},\cos\frac{k_z}{2}\cos\frac{k_x}{2}\right)$,   $\boldsymbol \Sigma=(-\Gamma^1,-\Gamma^3,\Gamma^{13})$, and the sum   is restricted to half Brillouin zone since  $\gamma^{\phantom\dagger}_{-\mathbf k}=\gamma^\dagger_{\mathbf k}$ \cite{Coleman1994}. As the components of $\boldsymbol \Sigma$ obey $[\Sigma^a,\Sigma^b]=i\epsilon^{abc}\Sigma^c$, the spectrum is given simply by \be
 \varepsilon_{\pm}(\mathbf k)=\pm |t||\mathbf h_{\mathbf k}|.
 \ee

 \begin{figure}[t]
\begin{center}
\includegraphics*[width=.99\columnwidth]{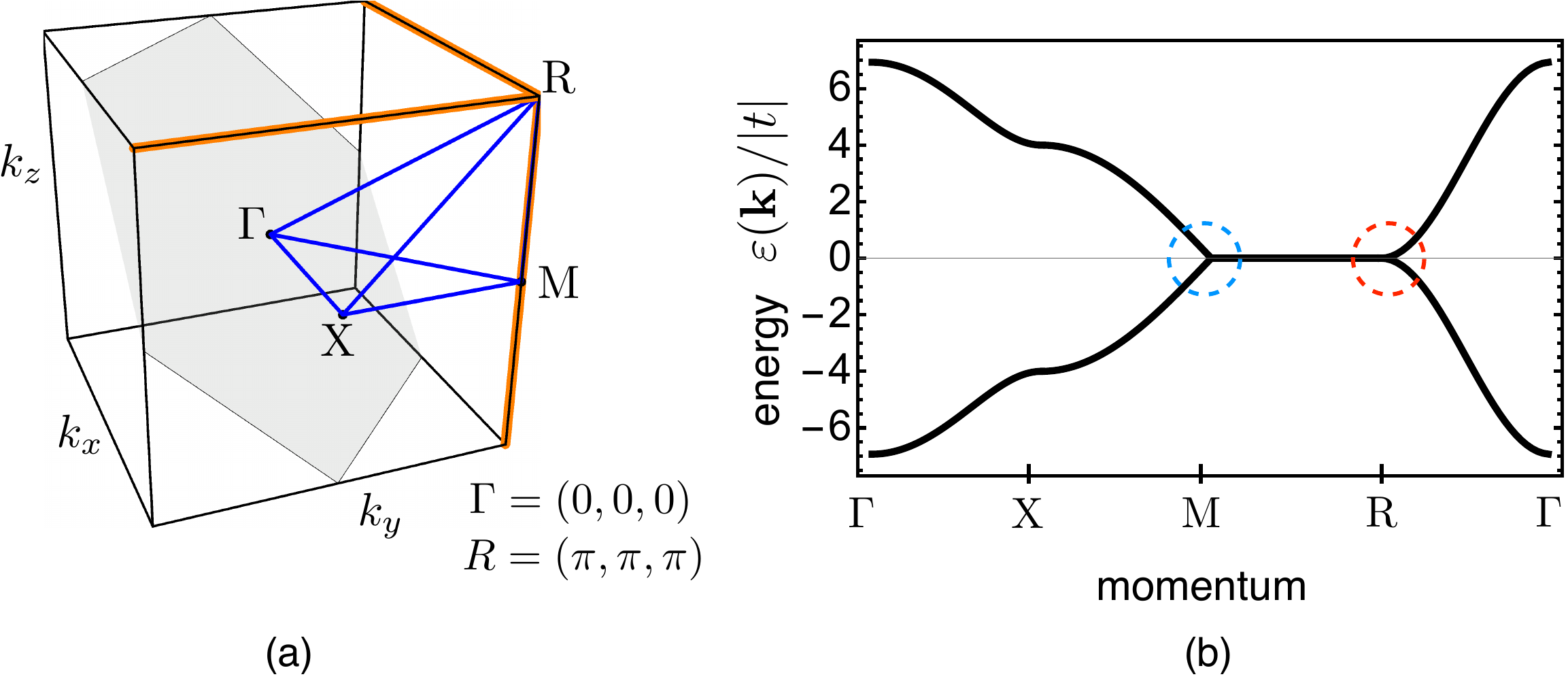}
\end{center}
\caption{\label{fig:Fermilines}(color online). (a) Brillouin zone of the cubic sub-lattice. (b) Bulk dispersion. The  nodal lines along    MR directions cross at the quadratic band touching point R. 
 }
\end{figure}

The dispersion relation is illustrated in Fig. \ref{fig:Fermilines}. There are two doubly degenerate  bands  \cite{Suppl}. Since $\{P,\tilde H_{\text{MF}}\}=0$, the Hamiltonian  has a chiral symmetry \cite{WenZee1989,Koshino2014} and the spectrum is symmetric between positive- and negative-energy states. The defining feature of the band structure is the band touching along the edges of the Brillouin zone. These are  nodal lines parametrized, e.g., by   $\mathbf k=(\pi,\pi,k_z)$. Expanding  $\mathbf k=  (\pi+p_x,\pi+p_y,k_z)$, with $p_x,p_y\ll 1$, we obtain the effective Hamiltonian on a plane perpendicular to a line node: $\mc H (\mathbf k)\approx 2|t|\cos\frac{k_z}2(p_x\Sigma^y+p_y\Sigma^z)$. The latter is formally equivalent to the Hamiltonian for graphene and yields   linear dispersion at low energies with $k_z$-dependent velocity $\varepsilon_\pm(\mathbf k)\approx \pm 2|t|\cos\frac{k_z}{2}\sqrt{p_x^2+p_y^2}$. 
These nodal lines can be characterized as  topological defects of an SU(2) Berry connection \cite{Wilczek1984} in reciprocal space (see Supplemental Material \cite{Suppl}).  
  The three nodal lines related by C$_3$ symmetry cross at $\text{R}=(\pi,\pi,\pi)$. Expanding $\mathbf k=  (\pi+p_x,\pi+p_y,\pi+p_z)$, we find that   R   is a quadratic band touching point \cite{Sun2009,Herbut2014} with anisotropic dispersion $\varepsilon_\pm(\mathbf k)\approx  |t|\sqrt{p_x^2p_y^2+p_y^2p_z^2+p_x^2p_z^2}$.

\begin{figure}[t]
\begin{center}
\includegraphics*[width=.65\columnwidth]{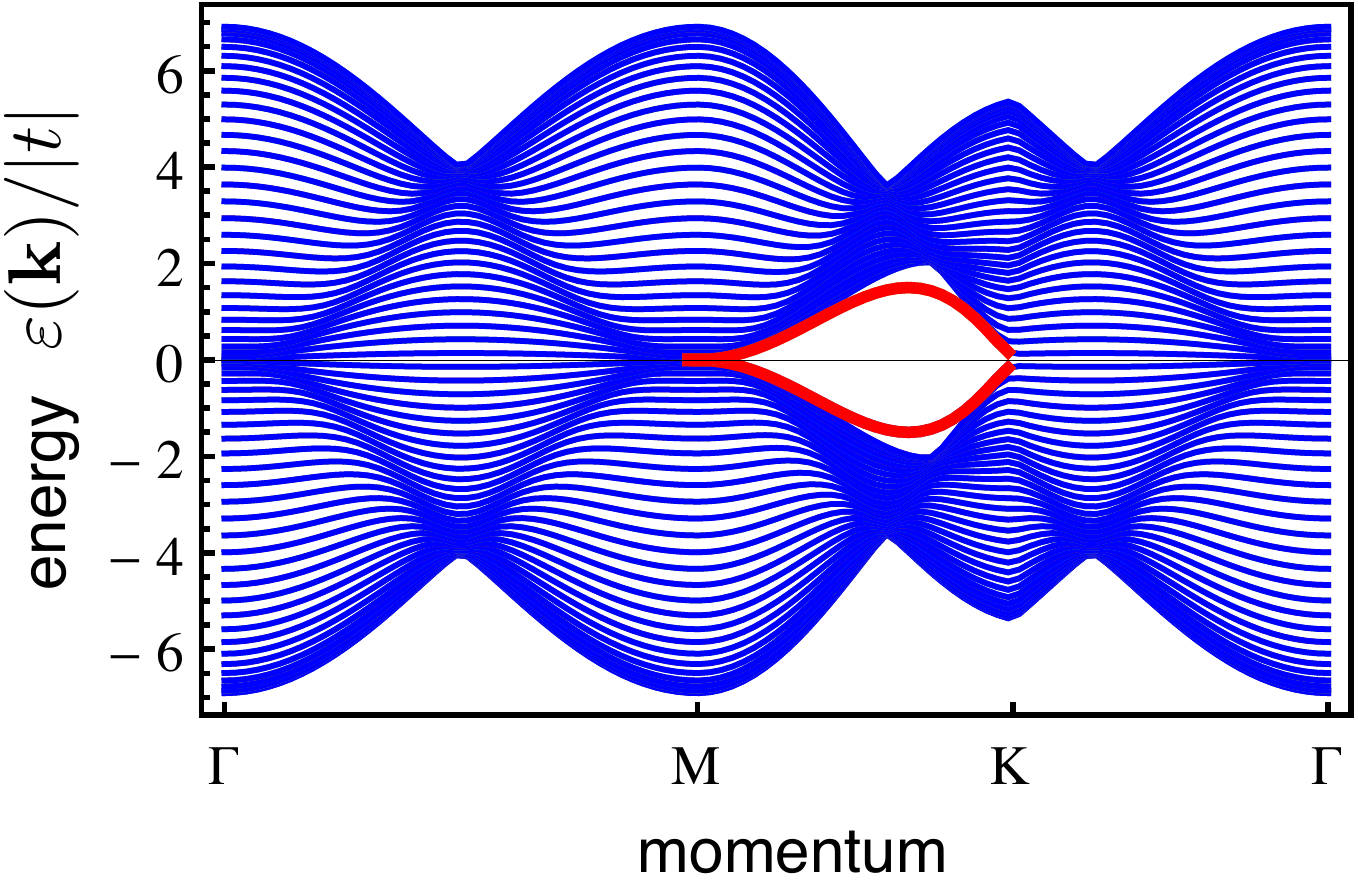}
\end{center}
\caption{\label{fig:surface}(color online). Surface-state spectrum projected in the Brillouin zone of the triangular lattice for (111) surface. The dispersion of the surface states corresponds to the thick red line between points M and K. 
 }
\end{figure}

Another feature of topologically nontrivial states of matter
is the presence of protected  surface states. We identify the surface states by calculating the spectrum for $\tilde H_{\text{MF}}$ in a slab geometry with open boundary conditions in the (111) direction (Fig. \ref{fig:surface}).  
There appear two pairs of  doubly degenerate   bands separated from the continuum, with  dispersion  terminating at the projections of the nodal lines. Remarkably,  the  positive-energy surface states are \emph{spatially separated} from the negative-energy ones as their wave functions are localized at opposite surfaces (which surface depends on the sign of the  hopping parameter). This is a direct manifestation of the breaking of inversion symmetry.

We calculate the ground state energy $E_{\text{gs}}$ at mean-field level by solving the self-consistent equations that determine the   order parameters in Eq. (\ref{eq:Hmf}). For this purpose we   had to diagonalize the Hamiltonian for coupled $\theta^1$ and $\theta^3$ fermions in Eq. (\ref{eq:Hmf}) and found that the spectrum again displays nodal lines   \cite{Suppl}. We obtain $E^{\text{MF}}_{\text{gs}}/NJ\approx -0.248$.  A better estimate of $E_{\text{gs}}$ can be obtained by implementing a Gutzwiller projection according to Eq. (\ref{eq:constraint}) using VMC \cite{Gros1989,Wang2009}. Considering a restrictive form of the wave function which neglects variations   in the population of the fermionic flavors   (see Supplemental Material \cite{Suppl}), we obtain  $E^{\text{VMC}}_{\text{gs}}/NJ=-0.40(1)$. This energy is already comparable to that of  the best  variational state identified in Ref. \cite{Chen2010}, namely  a valence bond solid  with $E_{\text{VBS}}/NJ\approx -0.417$. 
We expect the   spin liquid   to be stable   since small fluctuations of the Z$_2$ gauge field only induce weak short-range interactions \cite{Wen2004}, which are  irrelevant in the renormalization group sense for topological semimetals with point or line band touching in three dimensions \cite{Herbut2014}.  

The low-temperature  thermodynamic properties  
are governed by the   density of states  $\rho(\varepsilon)\propto \sqrt{\varepsilon}$ of the Majorana fermions, which is due to the quadratic band touching point. It follows that the QSL has heat capacity $C\propto T^{3/2}$, magnetic susceptibility $\chi \propto T^{1/2}$, and thermal conductivity $\kappa\propto T^{3/2}$ for  $k_BT\ll J$. Another important property is the correlation function $G(\mathbf r)={\langle \mathbf J_{j}\cdot \mathbf J_{j+\mathbf r}\rangle}$.  We find that $G(\mathbf r)$ vanishes when $\mathbf r$ connects sites on the same sublattice. For  vectors connecting different sublattices along (100) directions in the form $\mathbf r=\boldsymbol \delta+r\hat{\mathbf e}$, where $\boldsymbol \delta \in \{(\frac12,\frac12,0),(\frac12,0,\frac12),(0,\frac12,\frac12)\}$  and $\hat{\mathbf e}\in\{\hat x,\hat y,\hat z\}$,    the correlation decays at large distances as   $G(\mathbf r)\sim 1/r^4$. This power-law decay      coincides with the result for a Dirac point in two dimensions \cite{Wang2009}.

Finally, we address the comparison with available experimental results for the spin liquid candidate Ba$_2$YMoO$_6$. 
Aharen \emph{et al.} \cite{Aharen2010}
observed  that both the   heat capacity and the magnetic susceptibility
vanishes at low temperatures and have attributed this behavior to  a gapped collective spin singlet.  de Vries \emph{et al.} \cite{Vries2010}   proposed a picture of a valence bond glass, but noted that the muon spin relaxation is comparable to that of QSLs \cite{Vries2013}. 
%There is also evidence for  a heterogeneous state, signalling that disorder-driven ``dangling spins'' may play an important role. 
Here we propose that an alternative explanation for the vanishing heat capacity and susceptibility at low temperatures  is the vanishing density of states of our \emph{gapless} spin-orbital liquid with nodal lines.  A comprehensive study of the properties of this QSL in comparison with experimental results will be presented elsewhere \cite{Natori2016}.

To summarize, we have studied a realistic model for double perovskites in the regime of strong spin-orbit coupling. 
We proposed a new  spin liquid ansatz that gives rise to nodal lines in the spectrum of Majorana fermions. We   argued that  some experimental 
results for Ba$_2$YMoO$_6$ can be interpreted in terms of the vanishing  density of states predicted by our theory. 
We hope this work will  stimulate the search for strongly correlated materials hosting fractional excitations  
with  nontrivial  momentum-space topology \cite{Trebst2015,Maciejko2015}. 

\acknowledgements
This work was supported by Brazilian agencies FAPESP (W.M.H.N., E.C.A.) and   CNPq (E.M., R.G.P.).

\bibliographystyle{aipnum4-1}

\bibliography{article1}

\newpage
\onecolumngrid

\appendix
%\title{Supplemental Material}
%\maketitle
\section{Supplemental Material}
\appendix
\title{Supplemental Material for ``Chiral spin-orbital liquids with   nodal lines''}

\author{W. M. H. Natori}
\affiliation{Instituto de F\'{i}sica de S\~ao Carlos, Universidade de S\~ao Paulo, C.P. 369, S\~ao Carlos, SP,  13560-970, Brazil}
\author{E. C. Andrade}
\affiliation{Instituto de F\'{i}sica de S\~ao Carlos, Universidade de S\~ao Paulo, C.P. 369, S\~ao Carlos, SP,  13560-970, Brazil}
\affiliation{Instituto de F\'{i}sica Te\'{o}rica, Universidade Estadual Paulista,
Rua Dr. Bento Teobaldo Ferraz, 271 - Bloco II, 01140-070, S\~ao Paulo, SP, Brazil}
\author{E. Miranda}
\affiliation{Instituto de F\'{i}sica Gleb Wataghin, Unicamp, Rua S\'{e}rgio Buarque de Holanda, 777, CEP 13083-859 Campinas, SP, Brazil}
\author{R. G. Pereira}
\affiliation{Instituto de F\'{i}sica de S\~ao Carlos, Universidade de S\~ao Paulo, C.P. 369, S\~ao Carlos, SP,  13560-970, Brazil}

\maketitle
%\section{Supplemental Material}
%\setcounter{secnumdepth}{1}

\section{1. Symmetry of the spin liquid  ansatz}

Time reversal $T=Ke^{-i\pi J^y_{\text{tot}}}=K\prod_j(i\Gamma^{13}_j)$ acts on peudospins and pseudo-orbitals as \be
T^{-1}\mathbf s_jT=\mathbf s_j,\qquad T^{-1}\tau^{x,z}_jT=\tau^{x,z},\qquad T^{-1}\tau^{y}_jT=-\tau_j^{y}.
\ee
In the Majorana fermion representation for $\mathbf s$ and $\boldsymbol \tau$ this transformation can be implemented by $T=K\prod_j\theta^1_j\theta^3_j$. This is equivalent to complex conjugation combined with the operation $\theta^1_j\to -\theta_j^1$ and $\theta^3_j\to -\theta^3_j$.  Thus, if we focus on the  decoupled flavors $\gamma\in \{\eta^a,\theta^2\}$, we can take time reversal to be represented simply by complex conjugation. 

Let $\gamma_{j\ell}$ denote a  Majorana fermion  on site $j$ belonging to sublattice $\ell=\text{A,B,C,D}$. The operators $\gamma_{\mathbf k\ell}$ in momentum space are defined by\bea
\gamma_{j\ell}&=&\sqrt{\frac8N}\sum_{\mathbf k\in \frac12\text{BZ}}[\gamma^{\phantom\dagger}_{\mathbf k\ell} e^{i\mathbf k\cdot \mathbf R_{  j}}+\gamma^\dagger_{\mathbf k\ell} e^{-i\mathbf k\cdot \mathbf R_{  j}}],\label{modeexpansion}\\
\gamma_{\mathbf k\ell}&=&\sqrt{\frac{2}N}\sum_{j\in \ell}\gamma_{j\ell}e^{-i\mathbf k\cdot \mathbf R_j},
\eea
and $\gamma_{\mathbf k\ell}$ are normalized such that $\{\gamma^{\phantom\dagger}_{\mathbf k\ell},\gamma^\dagger_{\mathbf k^\prime\ell^\prime}\}=\delta_{\mathbf k\mathbf k^\prime}\delta_{\ell\ell^\prime}$. For each flavor of Majorana fermion we combine the four sublattice modes into a single ``spinor'' $\gamma_{\mathbf k}=(\gamma_{\mathbf k\text{A}},\gamma_{\mathbf k\text{B}},\gamma_{\mathbf k\text{C}},\gamma_{\mathbf k\text{D}})^t$.

In momentum space, time reversal takes $\mathbf k\to -\mathbf k$.
Up to a hopping  amplitude (determined by self-consistent equations, see next section), the mean-field Hamiltonian for a decoupled flavor is of the form $\tilde H_{\text{MF}}=\sum_{\mathbf k\in \frac12\text{BZ}}\gamma^\dagger_{\mathbf k}\mc H(\mathbf k)\gamma^{\phantom\dagger}_{\mathbf k}$ with \be
\mc H({\mathbf k})=i\left(\begin{array}{cccc}0&f(k_x,k_y)&f(k_y,k_z)&f(k_x,k_z)\\
-f(k_x,k_y)&0&-f(k_x,k_z)&f(k_y,k_z)\\
-f(k_y,k_z)&f(k_x,k_z)&0&-f(k_x,k_y)\\
-f(k_x,k_z)&-f(k_y,k_z)&f(k_x,k_y)&0\\
\end{array}\right),\label{4by4matrix}
\ee
where $f(k_a,k_b)=4\cos(k_a/2)\cos(k_b/2)$. Notice the factor of $i$. It follows that \be
T^{-1} \mc H(\mathbf k)T=\mc H^*(-\mathbf  k)=-\mc H(\mathbf k).\ee

We define inversion $P$ as the reflection by the mirror plane that   exchanges  A and C sublattices (plane perpendicular to $ \boldsymbol\delta_{yz}=(0,\frac12,\frac12)$). In momentum  space, $   P  :k_x\to k_x,k_y\to -k_z,k_z\to -k_y$. In addition, we have the action in the internal (sublattice) space   given by the matrix  (with determinant -1) \be
  P =\left(\begin{array}{cccc}0&0&1&0\\
0&1&0&0\\
1&0&0&0\\
0&0&0&1\end{array}\right).
\ee 
We also define the Z$_2$ gauge transformation that changes the sign of fermions on the B sublattice: \be
 G_P=\left(\begin{array}{cccc}1&0&0&0\\
0&-1&0&0\\
0&0&1&0\\
0&0&0&1\end{array}\right).
\ee
It is easy to check that inversion    anticommutes with the mean-field Hamiltonian:\be
(  P   G_P)^{-1}\mc H(P\mathbf k)  P   G_P=-\mc H(\mathbf k).
\ee
It follows that the combined transformation $PTG_P$  is a symmetry of the  Hamiltonian:\be
(  P   T  G_P)^{-1}  \mc H(P \mathbf k)  P   T G_P=\mc H(\mathbf k).
\ee

The C$_3$ rotation about a $(111)$ axis that leaves an A site invariant is represented by\be
  C_3=\left(\begin{array}{cccc}1&0&0&0\\
0&0&0&1\\
0&1&0&0\\
0&0&1&0\end{array}\right),
\ee
and the rotation in momentum  space takes $R_3:    k_x\to k_z,k_y\to k_x,k_z\to k_y$. In this case a gauge transformation is not required;  we obtain  immediately that \be
  C_3^{-1}   \mc H(R_3\mathbf k)    C_3=\mc H(\mathbf k).
\ee
 
The C$_2$ rotation along the $z$ axis that exchanges A$\leftrightarrow $B, C$\leftrightarrow$ D  is represented by 
\be
  C_2=\left(\begin{array}{cccc}0&1&0&0\\
1&0&0&0\\
0&0&0&1\\
0&0&1&0\end{array}\right),
\ee
and in momentum space $R_2:   k_x\to -k_x,k_y\to -k_y$.
We need to combine the C$_2$ rotation with the gauge transformation\be
  G_2=\left(\begin{array}{cccc}1&0&0&0\\
0&-1&0&0\\
0&0&1&0\\
0&0&0&-1\end{array}\right).
\ee
We then have \be
(  C_2  G_2)^{-1}\mc H(R_2 \mathbf k)(  C_2  G_2)=\mc H(\mathbf k).
\ee
Translation by $\boldsymbol \delta_{xy} =(\frac12,\frac12,0)$, which we denote by $  T_{xy}$, has the same effect of exchanging sublattices as the above  C$_2$ rotation. Thus, conjugation by $  T_{xy}  G_2$, together with $\mathbf R_j\to\mathbf R_j +\boldsymbol \delta_{xy}$ in real space,  is also a symmetry of the Hamiltonian   (and likewise for the equivalent translations in $yz$ and $xz$ planes).

Now consider a C$_4$ rotation along the $z$ axis going through an A site, which  exchanges C and D sublattices. This can be represented in sublattice space by\be
\hat  C_4=\left(\begin{array}{cccc}1&0&0&0\\
0&1&0&0\\
0&0&0&-1\\
0&0&1&0\end{array}\right).\label{C4matrix}
\ee
In momentum space, $R_4:k_x\to k_y,k_y\to-k_x$. Combining with the gauge transformation:\be
\hat G_4=\left(\begin{array}{cccc}1&0&0&0\\
0&-1&0&0\\
0&0&-1&0\\
0&0&0&1\end{array}\right),
\ee
we find \be(  C_4  G_4)^{-1}  \mc H(R_4\mathbf k)(  C_4  G_4)=-\mc H(\mathbf k).\ee
Thus, like $P$ and $T$, the C$_4$ rotation inverts the chirality of the ansatz.  It is then easy to see that $C_4G_4T $ is a symmetry of the Hamiltonian.

The C$_4$ rotation   can be used to construct a symmetry transformation that accounts for the twofold degeneracy of the Majorana fermion bands. 

It is also interesting to consider the C$_4$ rotation   that exchanges A and B sublattices, given by\be
\hat  C_4^\prime=\left(\begin{array}{cccc}0&-1&0&0\\
1&0&0&0\\
0&0&1&0\\
0&0&0&1\end{array}\right).\label{C4prime}
\ee
If we define this to be a rotation around $z$ axis in the opposite direction than the one in Eq. (\ref{C4matrix}), the transformation in momentum space is $R_4^\prime=(R_4)^{-1}$, i.e.,  $R_4^{\prime}:k_x\to -k_y,k_y\to k_x$. It is easy to check that  the composition $M=\hat  C_4\hat  C_4^\prime$  commutes with the Hamiltonian, $M^{-1}  \mc H( \mathbf k)M=\mc H(\mathbf k)$, and obeys  $M^2=-\mathbbm1$. Thus, we can block diagonalize $\mc H( \mathbf k)$ by sectors labeled by the eigenvalue $\pm i$ of the matrix $M$:\be
U_M^\dagger\mc H( \mathbf k)U_M^{\phantom\dagger}=\left(\begin{array}{cc}
\tilde{\boldsymbol\sigma}\cdot \mathbf h_{\mathbf k}&0\\
0&-\tilde{\boldsymbol\sigma}\cdot \mathbf h_{\mathbf k}
\end{array}\right),
\ee 
where $\tilde{\boldsymbol\sigma}\equiv (-\sigma^z,\sigma^y,\sigma^x)$ and $U_M$ is the unitary matrix that diagonalizes $M$. It is then clear that the spectrum of $\mc H( \mathbf k)$ is twofold degenerate with eigenvalues $\pm |\mathbf h_{\mathbf k}|$. Two degenerate states can be distinguished by the eigenvalue $\pm1$ of the Hermitean matrix $iM$ (which is analogous to the chirality of Weyl fermions in the massless Dirac equation).

In summary,  the chiral spin-orbital liquid ansatz lowers the symmetry of the Hamiltonian from O$_{\text h}\times$Z$_2$ (where   Z$_2$ is time reversal) to O$_{\text h}$ (where the new   group contains combinations of broken point group symmetries with the broken time reversal). 

\section{2. Berry connection}

The nodal lines can be characterized as  topological defects of a Berry connection in reciprocal  space. In our case, the Berry connection has to be  non-Abelian due to the double degeneracy of the bands. Away from the nodal lines, we   define the SU(2) connection   \begin{equation}
 A^a_{mn}(\textbf{k})=i\langle\psi_{m}(\textbf{k})|  \partial_{  k_a}\psi_{n}(\textbf{k})\rangle,
\end{equation}
where $|\psi_{m}(\textbf{k})\rangle$ and $|\psi_{n}(\textbf{k})\rangle$, $m,n\in\{1,2\}$, are degenerate eigenstates of $\mc H (\mathbf k)$ (say  with energy $\varepsilon_+(\mathbf k)$) chosen so as to obey $\langle \psi_m(\mathbf k)|\psi_n(\mathbf k)\rangle =\delta_{mn}$ and to diagonalize $A^z(\mathbf k)$. The generalized Berry phase is the Wilson loop \be
U=\mc P\exp[-i\oint dk_a A^a(\mathbf k)],\ee
where $\mc P$ denotes path ordering. The calculation of $U$ is simplified if we consider a  path around the line node parametrized by $\mathbf k\approx (\pi+\epsilon\cos\alpha,\pi+\epsilon \sin\alpha,k_z)$, with $\alpha\in [-\pi,\pi)$.   For  infinitesimal radius $\epsilon\ll 1$, we obtain \be
 A^x = -\frac{1}{2\epsilon }\sin \alpha\,\sigma^y+\mc O(\epsilon^0), \qquad A^y = \frac{1}{2\epsilon }\cos \alpha\,\sigma^y+\mc O(\epsilon^0),\ee
 which is precisely  the singular $\epsilon$ dependence   of a vortex line.  We then find $U=-\mathbbm 1$,  equivalent to a $\pi$ Berry phase.

\section{3. Solving the mean-field Hamiltonian}

In this section we outline the steps required to diagonalize the
mean-field Hamiltonian  and calculate  the ground state energy. 

Using the mode expansion Eq. (\ref{modeexpansion}), we can rewrite the various hopping terms for Majorana fermions in terms of operators in reciprocal space. For instance,\be
i\sum_{\langle j,l\rangle\in \text{XY}}\gamma_{j\text{A}}\gamma_{l\text{B}}=2i\sum_{\mathbf k\in\frac12\text{BZ}}h_{\mathbf k}^{1}(\gamma^\dagger_{\mathbf k \text A}\gamma^{\phantom\dagger}_{\mathbf k \text B}-\gamma^\dagger_{\mathbf k \text B}\gamma^{\phantom\dagger}_{\mathbf k \text A}), 
\ee 
where $h_{\mathbf k}^{1}=4\cos(k_x/2)\cos(k_y/2)$ is the first component of $\mathbf h_{\mathbf k}$. The mean-field Hamiltonian becomes
\bea
\tilde{H}_{\text{MF}} & =&\frac{J}{18}\sum_{\textbf{k}\in\frac{1}{2}\text{BZ}}\left[(2u+\bar{w})\sum_{a=1}^3\left(\eta^a_{\textbf{k}}\right)^{\dagger}\mc H_1(\mathbf{k})\eta_{\textbf{k}}^{a}-w\left(\theta_{\textbf{k}}^{2}\right)^{\dagger}\mc H_1(\mathbf k)\theta_{\textbf{k}}^{2}+\left(\Theta_{\textbf{k}}\right)^{\dagger}\mc H_2(\mathbf k)\Theta_{\mathbf k}\right]\nonumber\\
&&-\frac{NJ}{6}+\frac{J}{36}\underset{\langle i,j\rangle\in\alpha}{\sum}\left(3u_{ij}^{2}+3\bar{w}^{\alpha}_{ij}u_{ij}-w^{\alpha}_{ij}v_{ij}\right),\label{eq:Hmftotal}
\eea
where $\mc H_1(\mathbf k)$ is the $4\times4$ matrix in Eq. (\ref{4by4matrix}),  $\Theta_{\mathbf k}=(\theta^1_{\mathbf k\text A},\dots,\theta^3_{\mathbf k \text{D}})^t$ is an eight-component spinor that combines $\theta^1$ and $\theta^3$ fermions and  $\mc H_2(\mathbf k)$ is an $8\times8$ matrix to be specified  below.

First consider the fermions $\zeta\in \{\eta^a,\theta^2\}$, whose spectrum is determined by $\mc H_1(\mathbf k)$. Let  $U_{\textbf{k}}$ be the unitary transformation that diagonalizes
$\mc H_1(\mathbf k)$: 
\be
U^\dagger_{\mathbf k}\mc H_1(\mathbf k)U_{\mathbf k}^{\phantom\dagger}=\Lambda_1({\mathbf k}),
\ee
where $\Lambda_1({\mathbf k})=\text{diag}\{-|\mathbf h_{\mathbf k}|,-|\mathbf h_{\mathbf k}|,|\mathbf h_{\mathbf k}|,|\mathbf h_{\mathbf k}|\}$ is a diagonal matrix. The operators that annihilate fermions in eigenstates of $\mc H_1(\mathbf k)$ are\bea
\tilde \gamma_{\mathbf k\lambda}&=&\sum_{\ell=\text{A},\dots,\text{D}}(U^\dagger)_{\lambda\ell } \gamma_{\mathbf k\ell}, \\
  \gamma_{\mathbf k\ell}&=&\sum_{\lambda=1}^4(U)_{\ell \lambda}\tilde  \gamma_{\mathbf k\lambda}, 
\eea
where $\lambda=1,\dots,4$ is the band index. The mean-field  ground state $|\text{GS}\rangle $ is the state in which all single-fermion states with negative energy are occupied.  This leads to the self-consistent equation for  expectation values, e.g. \bea
\langle \gamma_{j,\text{A}}\gamma_{j+\boldsymbol{\delta}_{xy},\text{B}}\rangle&=&\frac{8}N\sum_{\mathbf k\in \frac12\text{BZ}}[e^{i\mathbf k\cdot  \boldsymbol{\delta}_{xy}}\langle  \gamma^\dagger_{\mathbf k\text{A}}\gamma^{\phantom\dagger}_{\mathbf k\text{B}}\rangle+e^{-i\mathbf k\cdot  \boldsymbol{\delta}_{xy}}\langle  \gamma^{\phantom\dagger}_{\mathbf k\text{A}}\gamma^{\dagger}_{\mathbf k\text{B}}\rangle]\nonumber\\
&=&i\text{Im}\left[\frac{16}N\sum_{\mathbf k\in \frac12\text{BZ}}\sum_{\lambda} (U^\dagger)_{\lambda\text{A} }U_{\text{B}\lambda }e^{i\mathbf k\cdot  \boldsymbol{\delta}_{xy}}\langle \tilde \gamma^\dagger_{\mathbf k\lambda}\tilde\gamma^{\phantom\dagger}_{\mathbf k\lambda}\rangle\right]\nonumber\\
&=&\frac{i}{2\pi^3}\text{Im}\left\{\sum_{\lambda\,(\epsilon_\lambda<0)}\int_{\frac12\text{BZ}} d^3k\,(U^\dagger)_{\lambda\text{A} }U_{\text{B}\lambda }\right\},\label{imagin}
\eea
where in the last line the sum is over bands with negative energy and we took the thermodynamic limit to replace $\sum_{\mathbf k}\to \frac{N}{32\pi^3}\int d^3k$ (corresponding to $N/4$ states in the Brillouin zone of the cubic sublattice).

Since $\mc H_1(\mathbf k)$ determines the spectrum of $\eta^a$ and $\theta^2$ fermions, the self-consistent equations for  $u_{ij}=-i\langle\eta_{i}^{a}\eta_{j}^{a}\rangle$ and
$v_{ij}=-i\langle\theta_{i}^{2}\theta_{j}^{2}\rangle$  are the same up to an overall minus sign,  depending on the relative  sign of the hopping amplitudes $2u+\bar w$ and $-w$ in Eq. (\ref{eq:Hmftotal}) We then have the constraint $|u|=|v|$, but must analyze two possibilities, namely $u=v$ and $u=-v$. Without loss of generality (by choosing one of the two degenerate ground states with opposite chiralities), we can set $u>0$. Numerical evaluation of the integral in Eq. (\ref{imagin}) then yields $u\approx 0.258$.

The relation between $u$ and $v$ determines the $8\times8$ matrix for $\Theta_{\mathbf k}$. For $v=u$, we obtain \begin{equation}
\mc H_2(\textbf{k})=u\textbf{h}_{\textbf{k}}\cdot\boldsymbol{\Sigma}',
\end{equation}
where $\boldsymbol{\Sigma}'=\left(2K^{z}\otimes\sigma^{y},-2K^{y}\otimes\mathbbm 1,-2K^{x}\otimes\sigma^{y}\right)$, with
\be
K^{x}  =\frac12\left(\begin{array}{cccc}
0 & 2 & 0 & \sqrt{3}\\
2 & 0 & \sqrt{3} & 0\\
0 & \sqrt{3} & 0 & 0\\
\sqrt{3} & 0 & 0 & 0
\end{array}\right),\quad 
K^{y}  =\frac{i}2\left(\begin{array}{cccc}
0 & -2 & 0 & \sqrt{3}\\
2 & 0 & -\sqrt{3} & 0\\
0 & \sqrt{3} & 0 & 0\\
-\sqrt{3} & 0 & 0 & 0
\end{array}\right),\quad
K^{z} =\frac12\left(\begin{array}{cccc}
1\\
 & -1\\
 &  & -3\\
 &  &  & 3
\end{array}\right).
\ee
The  components of the matrix vector $\mathbf K$  satisfy the SU(2) algebra. We then obtain the spectrum of  $\mc H_2(\mathbf k)$ and use it to solve the self-consistent equations for $w_{ij}=-i\langle \theta_i^\alpha\theta_j^\alpha\rangle$ and $\bar w_{ij}=-i\langle \bar\theta_i^\alpha\bar\theta_j^\alpha\rangle$ analogous to Eq. (\ref{imagin}). In this case of $u=v$ we find a self-consistent solution with $w\approx -0.081$ and $\bar w\approx 0.317$. Having fixed the order parameters, we obtain the mean-field ground states energy $E_{\text{MF}}(v=u)=\langle \tilde H_{\text{MF}}\rangle \approx -0.244 NJ$. 

For $v=-u$ we obtain 
\be
H_{\textbf{k}}^{(2)}=u\textbf{h}_{\textbf{k}}\cdot\boldsymbol{\Sigma}'',\label{eq:second Hk2}
\ee
where $\boldsymbol{\Sigma}''=\left(\left(2-\sigma^{xy}\right)\otimes\Sigma^{1},\left(2-\sigma^{yz}\right)\otimes\Sigma^{2},\left(2-\sigma^{xz}\right)\otimes\Sigma^{3}\right)$, with  $\boldsymbol \Sigma$ the matrix  vector defined in the main text, and the $2\times2$ matrices $\sigma^\alpha$ given by  $
\sigma^{xy}=\sigma^{z}$, $\sigma^{yz(xz)}=\frac{1}{2}(-\sigma^{z}\pm\sqrt{3}\sigma^{x})$. In this case we find a self-consistent solution with $w\approx 0.161$ and $\bar w\approx 0.318$.  The ground state energy is $E_{\text{MF}}(v=-u)\approx -0.248NJ$, slightly lower than the result for $u=v$. This is the value quoted in the main text. We note  that the small difference between the two energies may  change beyond the mean-field level. However, we have verified that  both solutions give rise to a spectrum  with  nodal lines along MR directions, qualitatively similar to the spectrum for $\eta^a$ and $\theta^2$ fermions. Therefore, the properties derived from the low-energy density of states $\rho(\epsilon)\propto  \sqrt{\epsilon}$ are generic.

\section{4. Variational Monte Carlo}
To check the viability of the proposed chiral spin-orbital liquid
beyond the mean-field level, we now enforce the local constraint exactly by
considering a Gutzwiller projection of the mean-field wave function
\citep{Gros1989} by means of a variational Monte Carlo calculation
\citep{Ceperley1977}. 

We begin by rewriting the Majorana fermions in terms of three Dirac
fermions, closely following the representation used in Ref. \citep{Wang2009}
\begin{equation}
c_{i}^{\dagger}=\frac{1}{2}\left(\eta_{i}^{2}+i\theta_{i}^{2}\right),\; d_{i}^{\dagger}=\frac{1}{2}\left(\eta_{i}^{1}+i\eta_{i}^{3}\right),\; f_{i}^{\dagger}=\frac{1}{2}\left(\theta_{i}^{1}+i\theta_{i}^{3}\right).\label{eq:dirac_def}
\end{equation}
In terms of this representation, the local constraint is now translated
into the fact that a given site may either have no Dirac fermions, a state
denoted by $\left|0\right\rangle $, or two Dirac fermions, in which case
there are three possible states at each site defined as
\begin{equation}
%\left|X\right\rangle =c^{\dagger}d^{\dagger}\left|0\right\rangle ,\;\left|Y\right\rangle =d^{\dagger}f^{\dagger}\left|0\right\rangle ,\;\left|Z\right\rangle =f^{\dagger}c^{\dagger}\left|0\right\rangle .\label{eq:xyz_def}
\left|x_i\right\rangle =c_i^{\dagger}d_i^{\dagger}\left|0\right\rangle ,\;\left|y_i\right\rangle =d_i^{\dagger}f_i^{\dagger}\left|0\right\rangle ,\;\left|z_i\right\rangle =f_i^{\dagger}c_i^{\dagger}\left|0\right\rangle .\label{eq:xyz_def}
\end{equation}
Given a real space configuration specified by the locations of the doubly occupied sites,
$X \equiv \left\{ x_{i}\right\},Y \equiv \left\{ y_{i}\right\},Z \equiv \left\{ z_{i}\right\}$, the wave function assigns
an amplitude $\Psi\left(\left\{ x_{i}\right\} ,\left\{ y_{j}\right\} ,\left\{ z_{m}\right\} \right)$
to it. Notice that the locations of the $\left|0\right\rangle $ states
are automatically specified. We point out that the local constraint
only fixes the parity of the number of fermions but not the number
itself. Moreover, our Hamiltonian contains terms which not only create/annihilate
two particles, e.g. $\left|x_i\right\rangle \leftrightarrows\left|0\right\rangle $,
but also terms which preserve the total number of fermions while
changing the number of each of the three individual fermionic flavors,
e.g. $\left|x_i\right\rangle \leftrightarrows\left|y_i\right\rangle $.
Although it is possible to write down a projected wave functions with
varying particle number \citep{Edegger2005}, we refrain from doing
so in this work for the sake of computational simplicity. Instead,
we consider a restrictive form for the ground state wave function:
each one of the four states is equally distributed over $N/4$ distinct
sites, with $\left\{ x_{i}\right\} =\left\{ x_{1},x_{2},\ldots,x_{N/4}\right\} $,
etc. 

The mean-field Hamiltonian in the main text may be rewritten, in terms
of the three Dirac fermions in Eq. \eqref{eq:dirac_def}, as 
\begin{eqnarray}
\mathcal{H}_{MF} & = & -\frac{NJ}{6}+\frac{J}{36}\sum_{\left\langle jl\right\rangle \in\alpha}\left[\left(3u_{jl}^{2}+3\bar{w}_{jl}^{\alpha}u_{jl}-w_{jl}^{\alpha}v_{jl}\right)+i\left(4u_{jl}+2\bar{w}_{jl}^{\alpha}\right)d_{j}^{\dagger}d_{l}\right.\nonumber \\
 & &+  \left.i\left(2u_{jl}+\bar{w}_{jl}^{\alpha}-w_{jl}^{\alpha}\right)c_{j}^{\dagger}c_{l}+i\left(2u_{jl}+\bar{w}_{jl}^{\alpha}+w_{jl}^{\alpha}\right)c_{j}^{\dagger}c_{l}^{\dagger}+i\left(3u_{jl}-v_{jl}\right)f_{j}^{\dagger}f_{l}+\mbox{h.c.}\right]\nonumber \\
 & &-  \frac{J}{36}\sum_{\left\langle jl\right\rangle \in XY}\left[i\left(3u_{jl}+v_{jl}\right)f_{j}^{\dagger}f_{l}^{\dagger}+\mbox{h.c.}\right]+\frac{J}{36}\sum_{\left\langle jl\right\rangle \in YZ}\left[i\left(3u_{jl}e^{i\pi/3}+v_{jl}e^{-2i\pi/3}\right)f_{j}^{\dagger}f_{l}^{\dagger}+\mbox{h.c.}\right]\nonumber \\
 & &+  \frac{J}{36}\sum_{\left\langle jl\right\rangle \in XZ}\left[i\left(3u_{jl}e^{i\pi/3}-v_{jl}e^{-2i\pi/3}\right)f_{j}^{\dagger}f_{l}^{\dagger}+\mbox{h.c.}\right].\label{eq:mf_dirac}
\end{eqnarray}
The three fermion flavors are decoupled and we may thus write the
mean-field wave function as a product of three Slater determinants. Thus, 
after the Gutzwiller projection, we obtain 
\begin{eqnarray}
\Psi\left(\left\{ x_{i}\right\} ,\left\{ y_{j}\right\} ,\left\{ z_{m}\right\} \right)= & \Phi_{d}\left(\left\{ x_{i}\right\} ,\left\{ y_{j}\right\} \right)\cdot\Phi_{f}\left(\left\{ y_{j}\right\} ,\left\{ z_{m}\right\} \right)\cdot & \Phi_{c}\left(\left\{ z_{m}\right\} ,\left\{ x_{i}\right\} \right).\label{eq:WF_guess}
\end{eqnarray}
The $d$-fermion sector of the Hamiltonian in Eq. \eqref{eq:mf_dirac} corresponds to
free fermions and thus their mean-field ground state is obtained
by filling up the states with negative energy. For the $c$ and $f$-fermions
we have a BCS-like Hamiltonian instead
and their ground state is given by the vacuum of their respective
Bogoliubov quasiparticles \citep{Gros1989}. The different status of the $d$ fermion is expected from symmetry: the hidden global SU(2) symmetry 
of the original Hamiltonian implies the global U(1) symmetry corresponding to the conservation of the total number of Dirac  fermions defined by a combination of $\eta$ Majorana fermions. 
On the other hand, there is no continuous symmetry associated with $\theta$ Majorana fermions; as a result, the total number of $c$ and $f$ fermions is not conserved. The need to work with BCS-type wave functions in our case should be contrasted with the case of SU(4) symmetric models \cite{Wang2009}, where the SU(4) symmetry implies the conservation of the  numbers of all three flavors of Dirac fermions.

After constructing the mean-field wave function, we then implemented
a variational Monte Carlo calculation of the Gutzwiller-projected ground state energy  $E^{\text{VMC}}_{\text{gs}}$.
  We started by generating an initial state in which we populate $N/4$ randomly chosen sites with the $x$-state $\left(\left\{ x_{i}\right\}\right)$, then $N/4$ of the remaining sites with the $y$-state $\left(\left\{ y_{i}\right\}\right)$, and finally $N/4$ of the further remaining sites with the $z$-state $\left(\left\{ z_{i}\right\}\right)$. Our Monte
Carlo moves consist in exchanging random pairs of sites containing
distinct states. We allow for moves involving widely separated sites --- and
which would not be connected by the Hamiltonian --- because this improves
the sampling over the space of configurations. We accept or reject
these moves according to the usual Metropolis algorithm. After $N$
such exchange attempts, we are said to have performed one Monte
Carlo sweep and after every sweep we compute $E^{\text{VMC}}_{\text{gs}}$. $N_{\mbox{warm}}$
sweeps are performed before measurements of physical quantities for
``thermalization''.  Averages are then performed over $N_{\mbox{sweep}}$
sweeps. We typically considered $N_{\mbox{warm}}=N_{\mbox{sweep}}\sim10^{4}$.
The results were obtained for lattices of size $N=4L^{3}$ with $L=4$,
$6$, and $8$. We find that the change in the ground state energy with $N$ is smaller than the Monte Carlo error bars for the system sizes considered here. Thus, we quote the results for $L=8$ as the converged ones.

We computed the ground state energy for the two sets of mean-field
parameters quoted in this supplemental material. For $u=v=0.258$,
$\bar{w}=0.317$, and $w=-0.081$ we obtain $E^{\text{VMC}}_{\text{gs}}=-0.39\left(1\right)NJ$.
As for $u=-v=0.258$, $\bar{w}=0.318$, and $w=0.161$ we obtain $E^{\text{VMC}}_{\text{gs}}=-0.40\left(1\right)NJ$.
Clearly, the Gutzwiller projection decreases significantly the mean-field
energy down to values which are already comparable to that obtained, for
instance, for a valence-bond covering of the lattice $\left(E_{\text{VBS}}=-0.417NJ\right)$
\citep{Chen2010}, thus showing that the proposed chiral spin-orbital
liquid is a competitive ground state candidate. 

In light of this favorable energy of our proposed ansatz, we conclude by
pointing out two important restrictions in our variational Monte Carlo
calculation that, once lifted, should further decrease the value of the
ground state energy $E^{\text{VMC}}_{\text{gs}}$:
\begin{enumerate}
\item For the quoted values of $E^{\text{VMC}}_{\text{gs}}$, we considered the optimal values
for the mean-field amplitudes $v$, $\bar{w}$ and $w$ obtained
\emph{before} the Gutzwiller projection, i.e., at the mean-field level;
\item The restrictive form of the considered wave function neglects
variations both in the populations of the fermionic flavors and in the
total number of fermions. 
\end{enumerate}
We stress that these restrictions were important for this first calculation
beyond mean field due to the complexity of the chiral spin-orbital
liquid ansatz considered here. We leave a more detailed investigation,
together with a more precise estimate for the variational energy of our spin liquid ansatz, for future work.

\end{document}